\documentclass[Final]{agujournal2019}

\usepackage{lineno,layouts}

\usepackage{sansmath} 
\sansmath

\usepackage{soul,hyperref}
\usepackage{natbib}

\hypersetup{
	breaklinks,
	colorlinks=true,
	linkcolor=Blue,
  urlcolor=Purple,
	citecolor=Purple,
 }

\usepackage[labelfont=bf]{caption}

\journalname{Journal of Advances in Modeling Earth Systems (JAMES)}

\begin{document}
	
	%
	%

	\title{Kinetic energy of eddy-like features from sea surface altimetry}
	
	%
	%
	
	
	

	\authors{Josu\'e Mart\'inez-Moreno\affil{1}, Andrew McC. Hogg\affil{1}, Andrew E. Kiss\affil{1},\\Navid C. Constantinou\affil{1}, and Adele K. Morrison\affil{1}} 
	
	\affiliation{1}{Research School of Earth Science and ARC Center of Excellence for Climate Extremes,\\Australian National University, Canberra, Australia}
	
	
	
	
	
	\correspondingauthor{Josu\'e Mart\'inez-Moreno}{josue.martinezmoreno@anu.edu.au}
	
	
	
	
	\vspace*{-0.5em}
	\begin{keypoints}
		\item Eddy kinetic energy is decomposed into the coherent mesoscale eddies part and the residual due to jets, waves, and large-scale variability. 
		\item The coherent eddy component of the eddy kinetic energy in the Southern Ocean has increased during the past two decades. 
		\item The coherent eddy field amplitude has increased, while the number of eddies has decreased over the last two decades in the Southern Ocean.
		
	\end{keypoints}

	\vspace*{-0.5em}
	
	%
	%
	
	\begin{abstract}
	    The mesoscale eddy field plays a key role in the mixing and transport of physical and biological properties and redistribute energy budgets in the ocean. Eddy kinetic energy is commonly defined as the  kinetic energy of the time-varying component of the velocity field.
	    However, this definition contains all processes that vary in time, including coherent mesoscale eddies, jets, waves, and large-scale motions. The focus of this paper is on the eddy kinetic energy contained in coherent mesoscale eddies. 
	    We present a new method to decompose eddy kinetic energy into oceanic processes.
	    The proposed method uses a new eddy-identification algorithm (TrackEddy). This algorithm is based on the premise that the sea level signature of a coherent eddy can be approximated as a Gaussian feature. The eddy Gaussian signature  then allows for the calculation of  kinetic energy of the eddy field through the geostrophic approximation. 
	    TrackEddy has been validated using synthetic sea surface height data, and then used to investigate trends of eddy kinetic energy in the Southern Ocean using Satellite Sea Surface Height anomaly (AVISO+). 
	    We detect an increasing trend of eddy kinetic energy associated with mesoscale eddies in the Southern Ocean. This trend is correlated with an increase of the coherent eddy amplitude and the strengthening of wind stress over the last two decades.
		
		\noindent\textbf{Plain summary}
		
		\noindent It is well accepted that climate change results in the intensification of the winds, in particular of those blowing over the Southern Ocean.
		Despite previous research showing an increase of the high-frequency motions in the Southern Ocean due to the intensification of the winds, we still do not know how swirling vortices of tens to hundreds of kilometers in the ocean have responded to climate change. 
		In this study, we use satellite observations of the sea surface height from 1993 to 2017 to look for changes in the swirling vortices. The focus of our study is on the Southern Ocean as it is one of the areas with more vortices and also plays a key role in controlling the climate.
		We find that the energy of the vortices has increased over the past two decades. Using our method, we are able to pinpoint that the energy increase occurs due to an increase in the mean amplitude of the vortices rather than in an increase in their number. Finally, the vortices show a clear response to the strengthening of winds in the Southern Ocean.
	\end{abstract}	
		
	\section{Introduction}
	
	Ocean variability is composed largely of mesoscale processes, which include coherent eddies, meandering jets and waves. These mesoscale processes mix and transport tracers such as heat, salt and biochemicals across ocean basins, and also redistribute momentum, potential vorticity and energy \citep{Wyrtki_Eddy_1976,Chelton_Global_2007,paper_zhang2014_oceanic,Foppert_Eddy_2017}. 
	However, the contribution of each mesoscale process to kinetic energy has not been fully explored, which is crucial to further understand the ocean circulation, ocean biology and to improve global ocean numerical models \citep{Farneti_The_2010,Beal_On_2011}.

	Kinetic Energy (KE) has been invoked as a measure to understand temporal and spatial oceanic variability \citep{White_Seasonal_1995,Kang_On_2017}. 
	KE is commonly divided into the Eulerian time-mean or Mean Kinetic Energy (MKE) and the time-varying or Eddy Kinetic Energy \citep{Robinson_Eddies_1983}. However, to avoid confusion between coherent eddies (noun) and time-varying processes commonly referred to in the literature as eddy (adjective). Here we will use the term Transient Kinetic Energy (TKE) to refer to the KE of the time-varying component:
	\begin{equation}
		\label{eq:ke}
		\underbrace{\overline{u^2 + v^2}}_{\overline{\textrm{KE}}} =  \underbrace{\bar{u}^2 + \bar{v}^2}_{\textrm{MKE}}  + \underbrace{\overline{u'^2} + \overline{v'^2}}_{\overline{\textrm{TKE}}},
	\end{equation}
	where $u$, $v$ correspond to the horizontal velocity components, $\bar{u}$, $\bar{v}$ the time-mean velocity components, and $u'$, $v'$ the time-varying velocity components.
	In many parts of the ocean, transient processes dominate the KE field, i.e., the TKE is more than an order of magnitude greater than the MKE \citep{Gill_Energy_1974}. 
	These regions include the Alaska Stream, Gulf Stream, Kuroshio Current, East Australian Current, Agulhas Current, and the Antarctic Circumpolar Current (ACC) \citep{Wyrtki_Eddy_1976,Richardson_Eddy_1983}.  These mesoscale-rich regions contain approximately $70\%$ of the global TKE, and it has been estimated that around $30\%$ of the global TKE can be attributed to mesoscale coherent eddy processes, as opposed to other transient mesoscale processes \citep{Chelton_The_2011}. This estimate includes the geostrophic velocities within eddy interiors. However, the Sea Surface Height (SSH) signature within the eddy boundaries is not only attributable to coherent eddies but may contain signatures from other mesoscale processes.
	
	The temporal evolution of mesoscale-rich regions located in the Southern Ocean (SO) indicates an increase in TKE over the last two decades \citep{Hogg_Recent_2015} due to the gradual increase of wind stress over the SO \citep{Swart_Observed_2012,Bracegirdle_Assessment_2013,Lin_Mean_2018,Young_Multiplatform_2019}. 
	Some studies suggest that the SO is in an ``eddy-saturated state", i.e., a state in which the time-mean transport is insensitive to the increase in winds and, therefore, the transient field readjusts to the wind. This hypothesis has been verified several times in numerical models, for example by \citet{ Hallberg_An_2001}, \citet{Meredith_Circumpolar_2006},  \citet{Nadeau_Influence_2012}, \citet{Marshall_Eddy_2017}, and \citet{Constantinou_eddysaturation_2019}, but only indications of it have been seen in observations \citep{Bning_The_2008,Firing_Vertical_2011,Chidichimo_Baroclinic_2014}. 
	
	It is well known in the literature that the surface transient field is highly coupled with the wind forcing \citep{Duhaut_Wind_2006,Hughes_Wind_2008,Byrne_Mesoscale_2016}.  Furthermore, \citet{Meredith_Circumpolar_2006} showed a lag of 2-3 years between the area-averaged TKE and the circumpolar wind stress anomaly. This result was further confirmed regionally using numerical models in the SO \citep{Morrow_Eddy_2010} and the ACC \citep{Patara_Variability_2016}. All studies discussed thus far include all transient processes in their mean-transient decomposition; thus, the signature of just the coherent mesoscale eddies to the TKE trends remains unclear. 
	
	Oceanic coherent eddies have been studied through a variety of detection and tracking algorithms, mostly using either diagnostic methods or analytical methods.
	Diagnostic methods build on physical intuition to categorize coherent features of the flow based on physical and geometrical criteria. These methods are mostly based on automated eddy detection algorithms. One of the first studies relied on a measure of rotation and deformation known as the Okubo-Weiss parameter \citep{Chelton_Global_2007}. 
	However, the Okubo-Weiss approach has been criticized for its dependence on thresholds and its sensitivity to noise \citep{Chelton_The_2011,Souza_Comparison_2011}. 
	More recent methods include analysis based on wavelets \citep{Turiel_Wavelet_2007}, reversal of the flow field \citep{Nencioli_A_2010}, perturbation of the sea surface temperature \citep{Dong_An_2011}, the outermost closed Sea Surface Height anomaly (SSHa) contours \citep{Chelton_The_2011}, or a combination of physical and geometric parameters \citep{Viikme_Quantification_2013}, single extreme Sea Level Anomaly (SLA) contours \citep{Faghmous_A_2015}, and machine learning using the phase angle between velocity components \citep{Ashkezari_Oceanic_2016}. Analytical methods define eddies as coherent structures by mathematical estimations of coherence. Some of these studies include Lagrangian coherent structures identified by material rotation relative to the mean rotation of the deforming fluid volume \citep{Haller_Defining_2016,Tarshish_Identifying_2018}, the change in location of a fluid particle  induced by infinitesimal changes in its initial position (finite-time Lyapunov exponent) \citep{BeronVera_Oceanic_2008,Hadjighasem_A_2017}, and geometrical analysis using transfer operators and invariant manifolds \citep{Froyland_Detection_2007,Froyland_Almost_2009}.
	
	In this study, we present TrackEddy, a diagnostic method for eddy tracking. The main objective of TrackEddy is to capture the full coherent eddy field influence instead of only the material core (analytical method). The novelty of this algorithm is  its capability to reconstruct the mesoscale eddy field from global SSHa by fitting optimal anisotropic Gaussians to each identified eddy (first described by \citet{McWilliams_Anisotropic_1994}). Then the reconstructed field can be used to extract the kinetic energy contained in the coherent eddy field through the geostrophic approximation. 
	This Python open-source software builds on the algorithms developed by \citet{Fernandes_Discovery_2006}, \citet{Chelton_The_2011}, \citet{Viikme_Quantification_2013}, and  \citet{Faghmous_A_2015} and it is available at \url{https://github.com/josuemtzmo/trackeddy}. The new tracking-reconstruction algorithm and kinetic energy decomposition are detailed in section 2. 
	These methods have been tested using ensembles of synthetic data (section~3). 
	The analysis and results from the AVISO+ dataset (section~4) include a quantitative validation of the method, an update of the Transient Kinetic Energy trend associated only with eddy-like features in the SO and the response of eddies to the westerly wind intensification. 
	Our goal is to use these results to investigate whether the eddy field has a direct response to the wind intensification.

	\section{Methods}
	
	TrackEddy is an autonomous eddy identification, tracking, and reconstruction algorithm, which assumes eddies can be represented as isolated anisotropic Gaussian anomalies. 
	The main and unique characteristic of the TrackEddy algorithm, which differs from previous algorithms \citep{Chelton_Global_2007,Faghmous_A_2015,Ashkezari_Oceanic_2016}, is its capability to reconstruct an optimal Gaussian anomaly for each identified eddy. This Gaussian anomaly can be used to reconstruct the eddy velocities to calculate the TKE associated with the identified coherent eddies.
	
	TrackEddy follows a similar work-flow to previous methods using SSH. 
	It starts with a single snapshot of SSHa, where potential eddies are isolated using study-specific criteria. 
	Generally, each study describes a strict definition of what will be considered an eddy, by constraining their size and/or shape. 
	Then, the algorithm iterates at multiple discrete SSHa levels in which the coherent eddy definition is used to identify eddies. The identification algorithm at each discrete SSHa level is then applied to all time-steps for which data is available.
	The following subsections present the TrackEddy algorithm structure, criteria, user-specified values, and energy calculation.
	
	\subsection{Eddy Identification}
		
	TrackEddy starts at the extremum contour of the SSHa field, which corresponds to the maximum value or minimum value of the field anomaly. 
	Then, closed contours are identified and extracted for each contour level defined by the user. The finer the discrete step between contours, the more accurate the eddy sizes and the better the optimal Gaussian fit will be. 
	To be identified as a potential eddy, each closed contour must satisfy three main criteria. First, as \citet{Fernandes_Discovery_2006} proposed, eddies can be identified by using the optimal fitted ellipse. In the case of TrackEddy, the Pearson correlation coefficient of an optimal fitted ellipse, should be less than $R_{\epsilon}$, where the default value of $R_{\epsilon}$ is 0.9. 
	Second, the eccentricity defined as  $e=\sqrt{1-b^2/a^2}$, where $a$ corresponds to the major axis and $b$ to the minor axis of the ellipse should be greater than a threshold value $e_c$, which we defined as 0.85. This corresponds to a ratio of $a/b$ about $\sim 2$.
	Third, the area of each potential eddy contour should be smaller than $4 \pi^2 L_D^2$  \citep{Klocker_Global_2014}, where $L_D$ is the first-baroclinic Rossby radius of deformation taken from \citet{Chelton_Geographical_1998}.
	When these three criteria are met, the optimal Gaussian is fitted. 
	To constrain this optimization, the Gaussian amplitude and location are fixed to the maximum SSHa value inside the closed contour and the coordinates of this maximum, respectively. 
	The Gaussian spread and orientation are then optimized to obtain the best anisotropic representation of the eddy signature. 
	To ensure the best representation of the eddy field, each fitted Gaussian is tested by comparing the absolute difference between the integrals of the original field and the optimal fitted field. 
	If the absolute difference between the fields is larger than 10 percent of its original value the closed contour is discarded. 
	Finally, this process is repeated for each SSHa discrete level and for each time-step of the dataset. From all the Gaussian candidate fits for a single eddy at each time-step, TrackEddy only records the one where the integral of the Gaussian fit agrees the best with the integral of the SSHa field within the closed contour.
	
	The above-mentioned criteria mostly identify eddies with a single extreme value in each closed contour, but it is possible to identify multiple extrema in different contour levels when eddies merge and/or interact with other features. 
	There are additional sanity criteria which remove eddy candidates if the SSHa profiles over the minor and major axis of the fitted ellipse do not approximate a Gaussian, or if features are mostly surrounded by land. For the eddy identifications from the SSH fields in this study we verified that these additional criteria discarded less than $1\%$ of the identified eddies, however they are crucial to avoid unrealistic Gaussian fits. For more details on the TrackEddy algorithm, the reader is referred to the online documentation at \url{https://trackeddy.readthedocs.io}.
	
\begin{figure}[!t]
		\centering
		\includegraphics[width=0.5\textwidth]{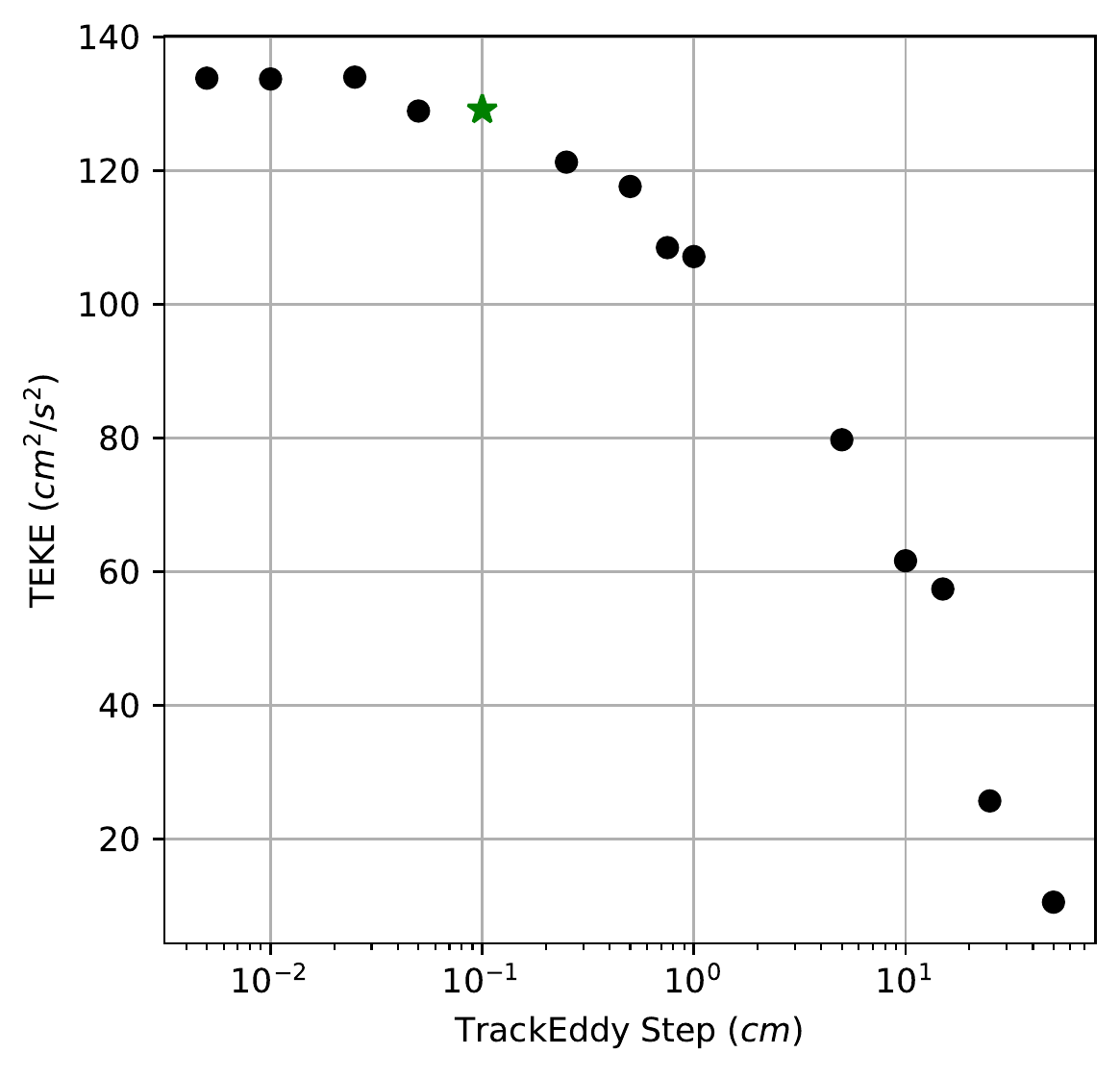}
		\caption{Convergence of TEKE computed from TrackEddy reconstruction by varying the identification step from 50~$cm$ to 0.005~$cm$ over the Agulhas region. Green star corresponds to the selected identification step for the analysis presented in section 4.}
		\label{fig:steps_time}
	\end{figure}

	The eccentricity parameter space $e_c$ was explored from 0.5 to 0.95 in steps of 0.15.  When the eccentricity value was 0.5, only coherent eddies with neglectable anisotropy were identified, while using eccentricity of 0.95 the algorithm started fitting features that could be identify as meanders. The fitting ellipse parameter $R_{\epsilon}$ was not explored as it only ensures the optimal ellipse to fit the eddy closed contour.
	Additionally, the best qualitative eddy reconstruction from the AVISO+ satellite dataset was produced using the values for $R_{\epsilon} = 0.9$ and $e_c=0.85$. A crucial parameter on the coherent eddy identification is the step in which closed contours are analyzed. We defined these steps as $0.1\ cm$ (green star in figure \ref{fig:steps_time}), where the TEKE over the Agulhas regions started converging approximately to $133\ cm^2/s^2$ (computational wall-time increases linearly with the number of steps). These parameters are used in the application of TrackEddy to the synthetic and satellite data presented in section~4. 
	
	\subsection{Kinetic energy decomposition}
	Kinetic energy is commonly separated into the mean and transient components by a Reynolds decomposition. At a given time, the velocities  ($u,v$) are split into their time-mean ($\bar{u},\bar{v}$) and time-varying components ($u'=u-\bar{u}$,\quad $v'=v-\bar{v}$).
	We further spatially decompose the time-varying velocities ($u'$, $v'$) into, e.g.,
	\begin{equation}
		u' = u'_e+u'_r,
		\label{eq:vel}
	\end{equation}
	and similarly for $v'$. In Eq.~(\ref{eq:vel}), $u'_e$  is the coherent eddy velocity defined as the geostrophic velocity computed from the optimal Gaussian fit and $u'_r$, the residual velocity is the difference between the geostrophic transient velocity and the coherent eddy  geostrophic velocity.
	Based on this velocity decomposition, TKE can be written as: 
	\begin{equation}
		\label{eq:tke}
		\underbrace{{u'}^2 + {v'}^2}_{\textrm{TKE}} =  \underbrace{{u'_{e}}^2 + {v'_{e}}^2}_{\textrm{TEKE}}  + \underbrace{{u'_{r}}^2 + {v'_{r}}^2}_{\textrm{TRKE}}   + \underbrace{2({u'_{e}}{u'_{r}} + {v'_{e}}{v'_{r}})}_{\textrm{TRKE}_c},
	\end{equation}
	where the TEKE term contains only energy from coherent eddy processes, TRKE is the energy computed from the geostrophic velocities of the non-coherent processes, and $\textrm{TRKE}_c$ are cross terms or the overlap between the coherent eddy field and the residual. 
	
	\section{Algorithm validation}
    
\begin{figure}[!t]
    	\centering
    	\includegraphics[width=0.8\textwidth]{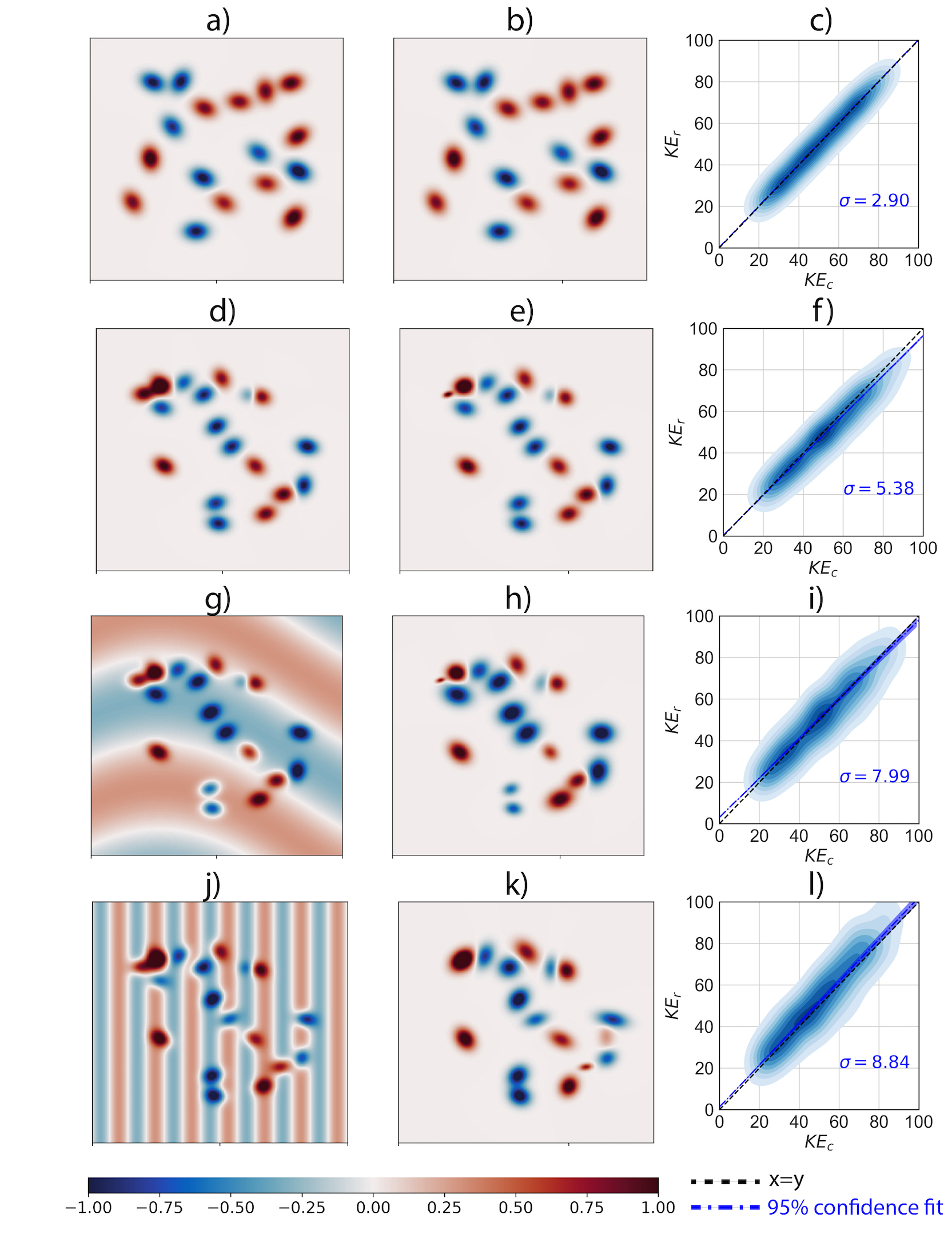}
    	\caption{Field plots show a single member of each synthesized SSH dataset ensembles and its reconstruction by the TrackEddy algorithm;
    		(a-b) no interaction control, (d-e) interaction control, (g-h) interacting eddies and propagating waves, and (j-k) interacting eddies and jets. Additionally, the 1000 members density distribution of the integrated control field KE ($\textrm{KE}_c$) versus the integrated reconstructed field kinetic energy ($\textrm{KE}_r$) correspond to panels c, f, i, and l. 
    	}
    	\label{fig:E_energy_density}
    \end{figure}

    We evaluate the quality of identified features by testing the algorithm with four ensembles of synthetic fields, each with 1000 members, created by the addition of randomly distributed Gaussian features. Each member contained a random number of Gaussian perturbations ($5<n<20$) at random locations with normal-distributed random properties. Each Gaussian has a polarity of $-1$ or $1$, an orientation from 0 to 180$^\circ$, and an amplitude and a major axis between 0.7 and 1.3. 
    The first and simplest experiment is a set of randomly distributed Gaussians constrained so that they do not overlap with any other Gaussian feature within a circle with radius of their major axis (no interaction control). Figure~\ref{fig:E_energy_density}a is an example of a single member with 17 Gaussian features of varying size. Figure~\ref{fig:E_energy_density}b shows the reconstruction of the features verifying that they have the correct location and the right Gaussian spread and orientation. The domain integrated KE of the non-interacting control and the reconstruction is shown in figure~\ref{fig:E_energy_density}c. 
    Therefore, TrackEddy can estimate the energy contained by non-interacting isolated Gaussians, that represent non-interacting eddies.
    
    Non-interacting eddies are a simple idealization of the ocean eddy field. We now consider progressively less idealized cases, beginning with interacting Gaussian features (interacting control). The second ensemble allows overlapping between Gaussians, which produces complex structures, such as the generation of elongated features when two or more Gaussians partially overlap, or large slopes when Gaussians of different polarity overlap. figures~\ref{fig:E_energy_density}d~\&~e show a sample member and its reconstruction from the interacting control ensemble. When Gaussians with opposite polarities partially overlap, the algorithm is able to identify and reconstruct the features. In the case of Gaussians with the same polarity, if each feature has an identifiable maximum, then the algorithm will fit the corresponding number of Gaussians shown in figures~\ref{fig:E_energy_density}d~\&~e. However, almost complete overlaps with identifiable independent closed contours, containing minimal information to optimize a Gaussian fit, will be represented poorly (figure~\ref{fig:E_energy_density}e). The integrated KE of the interacting control against the reconstruction shows a good estimation and the spread of the distribution about the one to one diagonal has  standard deviation $\sigma=5.38$, which is larger than the standard deviation of the non-interacting experiment ($\sigma=2.90$). We do not expect every feature to be perfectly reconstructed, particularly when the eddy-like features overlap. However, TrackEddy is able to identify and reconstruct the majority of features, and thereby represent the eddy signature and their kinetic energy content.
    
    To attempt more ``realistic'' evaluations, the remaining experiments use the same field as the interaction control experiment, but with background perturbations like waves (figure~\ref{fig:E_energy_density}g) and jets (figure~\ref{fig:E_energy_density}j). The experiment with waves is analogous to the interacting control, where most eddies are identified except when the Gaussians overlap almost completely. Strongly interacting Gaussians are still poorly represented (figure~\ref{fig:E_energy_density}h). Note that the amplitude of the reconstructed Gaussians depends on whether the background anomaly has the same or opposite sign. Thus, a larger spread of the standard deviation ($\sigma = 7.99$) is generated when comparing the reconstructed KE and the interacting control energy. Furthermore, when the jet-like background field (in which the sinusoidal pattern has a length-scale similar to the Gaussian) is used, most of the features are identified. However, there are some false positive identifications as shown in figure~\ref{fig:E_energy_density}k and an even larger standard deviation ($\sigma = 8.84$). Despite the misreadings in amplitude and number of features for both background perturbations, figures~\ref{fig:E_energy_density}i \& l show the reconstruction of KE using the TrackEddy algorithm approximates the energy contained by the control experiments. 
    
    In each of the evaluation tests shown here, the overall reconstruction of KE using the TrackEddy algorithm approximates well the energy contained by the control experiments. Therefore, we conclude that our algorithm is capable of representing and extracting the energy, even when there is a background perturbation field. In the next section, we proceed to use TrackEddy to reconstruct the eddy field and energy from the satellite SSHa field.
    
	\section{Results}
	
	After testing the capabilities of TrackEddy, we applied the algorithm to the global gridded AVISO+ satellite SSH product derived from all the available satellites from CMEMS (E.U. Copernicus Marine Service Information). 
	The daily analyzed period covers from January 1993 to December 2017 on a 0.25$^\circ$ $\times$ 0.25$^\circ$ longitude-latitude grid. However, the effective resolution of AVISO+ is coarser than 0.25$^\circ$ degrees, and therefore the capability to identify small scale coherent eddies is limited \citep{Amores_Up_2018}.
	SSHa data was obtained by removing the historical SSH climatology from 1993 to 2012 for each individual SSH snapshot and also removing the moving average of a 20$^\circ$ latitude/longitude kernel to preserve only mesoscale features. 
	The analysis and post-processing of the satellite data were parallelized in time (21 day chunks) using 448 cores. The implementation of TrackEddy in the supercomputer Raijin took approximately 67 hours (wall-time) or 13,000 hours in a single core to analyze the presented results.
	The global eddy database identified using TrackEddy from the satellite AVISO+ dataset is publicly available (refer to Acknowledgements for dataset DOI).
	
	\subsection{Transient Kinetic Energy}
	
\begin{figure}[!t]
	\centering
	\includegraphics[width=1\textwidth]{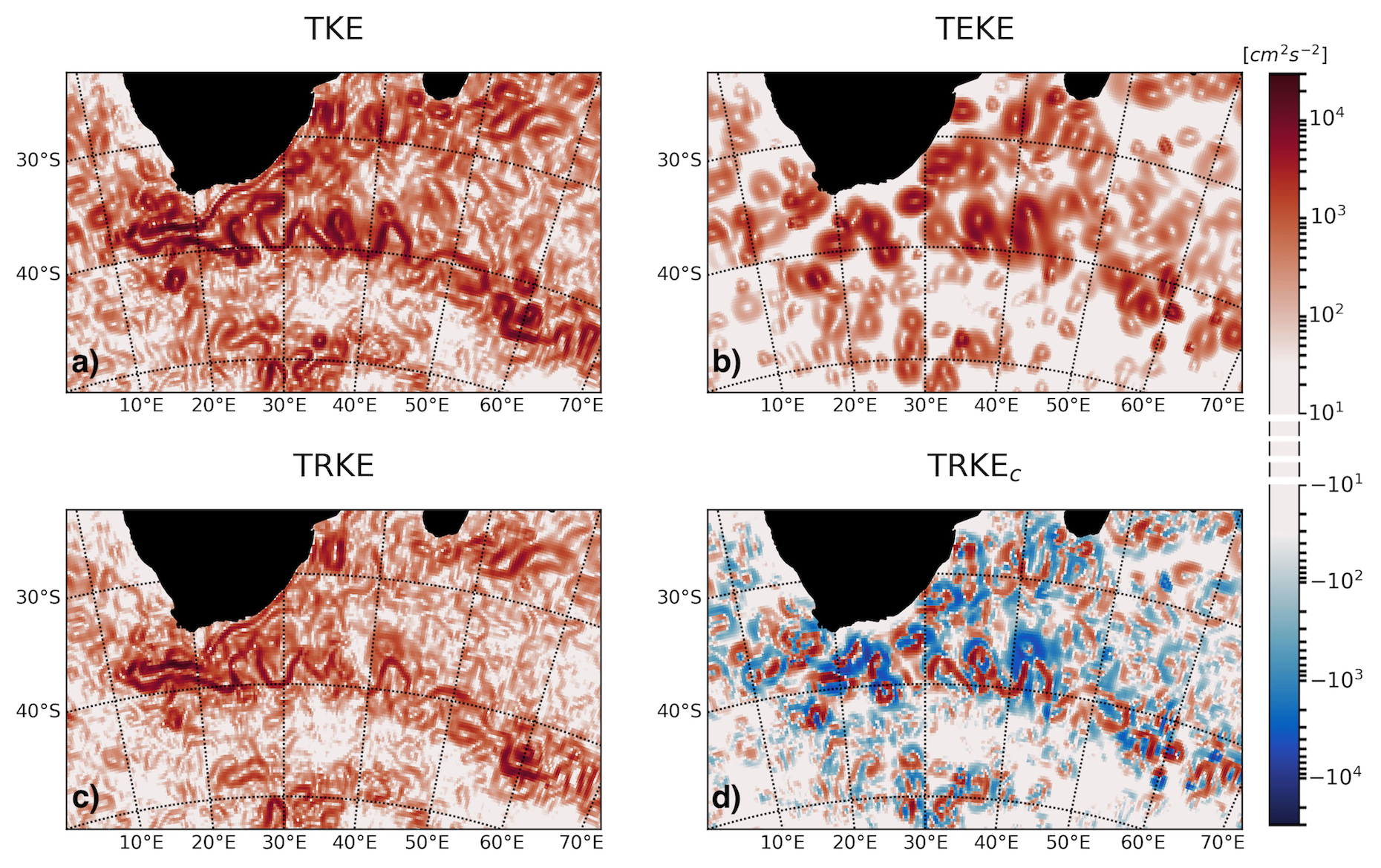}
	\caption{Magnitude of Transient Kinetic Energy and its decomposition in the Agulhas Current  for a snapshot on January 1st 2016.
		a) Transient Kinetic Energy, b) Transient Eddy Kinetic Energy or the energy of eddy processes, c) the Transient Residual Kinetic Energy or energy of jets and waves, and d) the cross terms which correspond to the overlap between processes.}
	\label{fig:agulhas_zoom}
	\end{figure}

	The proposed transient kinetic energy decomposition contains the energy from coherent eddy processes (TEKE), non-coherent processes (TRKE), and cross terms between the coherent eddies and non-coherent processes  ($\textrm{TRKE}_c$). Figure~\ref{fig:agulhas_zoom} shows a snapshot from January 1st 2016 of the TKE, TEKE, TRKE and $\textrm{TRKE}_c$ fields in the Agulhas Current region. Figure~\ref{fig:agulhas_zoom}a shows a TKE snapshot where ring-like features and filaments can be observed, corresponding to coherent eddies, and jets respectively. The signature of coherent eddies in Kinetic Energy or TEKE in the Agulhas region is shown in figure~\ref{fig:agulhas_zoom}b using an shape preserving projection (Lambert Conformal). This snapshot shows elliptical areas with large KE values. Each individual eddy is shown as a ring with two local maxima on either side of the major axis (figure~\ref{fig:agulhas_zoom}b). These local maxima result from the elliptical nature of the reconstructed eddies. 
	
\begin{figure}[!t]
		\centering
		\includegraphics[width=1\textwidth]{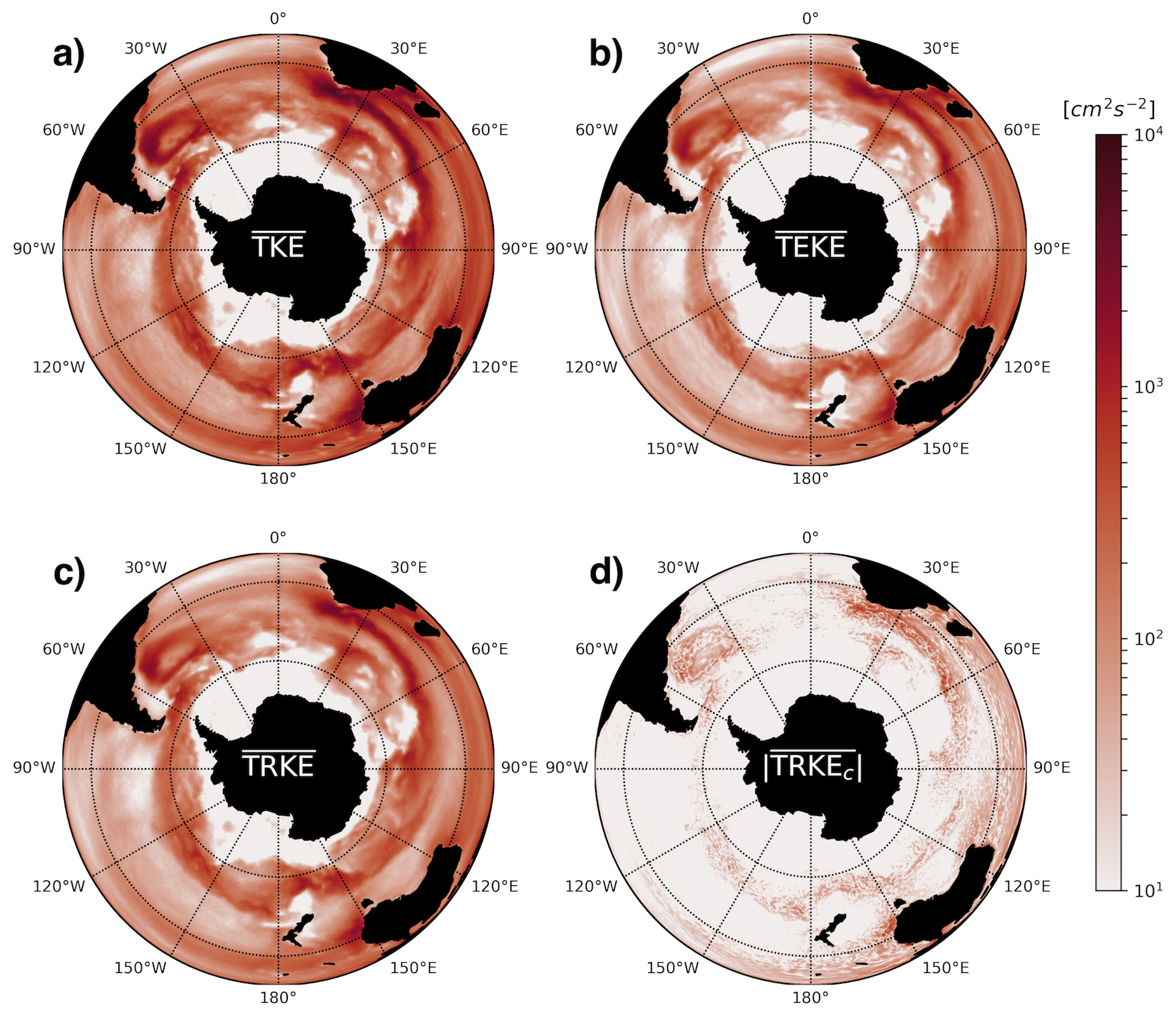}
		\caption{Magnitude of average Transient Kinetic Energy and its decomposition in the Southern Ocean reconstructed from satellite SSHa from 1993 to 2017. 	a) Transient Kinetic Energy, b) Transient Eddy Kinetic Energy or the energy of eddy processes, c) the Transient Residual Kinetic Energy or energy of jets and waves, and d) the cross terms which correspond to the overlap between processes.}
		\label{fig:components_diagnostic}
	\end{figure}

	The Southern Ocean time-mean values of TKE and TEKE are shown in figure~\ref{fig:components_diagnostic}a\&b respectively. The mean TKE (figure~\ref{fig:components_diagnostic}a) is several orders of magnitude larger at the western boundary currents and the ACC than any other region in the SO. The mean TEKE (figure~\ref{fig:components_diagnostic}b) also shows the pathways of the ACC and the western boundary currents, which are key in the generation of coherent eddies. Finally, TEKE is fundamental to the understanding of the TKE as on average it explains $41.6\%$ of TKE in the Southern Ocean (0$^\circ$E - 360$^\circ$E, 30$^\circ$S - 60$^\circ$S) with a temporal variability of $9\%$, similar to the global estimate proposed by \citet{Chelton_The_2011}.
	
	The TRKE snapshot (figure~\ref{fig:agulhas_zoom}c) shows filaments which mostly correspond to jets, while some ring-like features are still observable, corresponding to eddies missed or imperfectly fitted by our algorithm. Again, the largest signatures in figure~\ref{fig:components_diagnostic}c are located in the ACC and western boundary currents. The mean TRKE (figure~\ref{fig:components_diagnostic}c) now mostly consists of jets, meanders, and waves. These processes contain approximately $57.7 \pm 9\%$ of the TKE in the Southern Ocean. Finally, the Gaussian fit may misrepresent the eddy signature, for example, if kurtosis is present, the split signal  between the eddy and the residual velocities (Eq. \ref{eq:tke}) will be collocated, which will result in the cross terms ($\textrm{TRKE}_c$) shown in figure~\ref{fig:components_diagnostic}d. The absolute magnitude of this field is smaller than TEKE and the structure mostly contains a ``random" spatial distribution of positive and negative values (figure~\ref{fig:agulhas_zoom}d). Therefore, $\overline{\textrm{TRKE}_c}$ is much smaller than any of the other components (figure~\ref{fig:components_diagnostic}), where the average signature of $|\overline{\textrm{TRKE}_c}|$ over TKE is  $1.3 \pm 0.6\%$, so $\textrm{TRKE}_c$ will be neglected as it is two order of magnitude smaller than the other components.
	
\begin{figure}[!t]
	\includegraphics[width=1\linewidth]{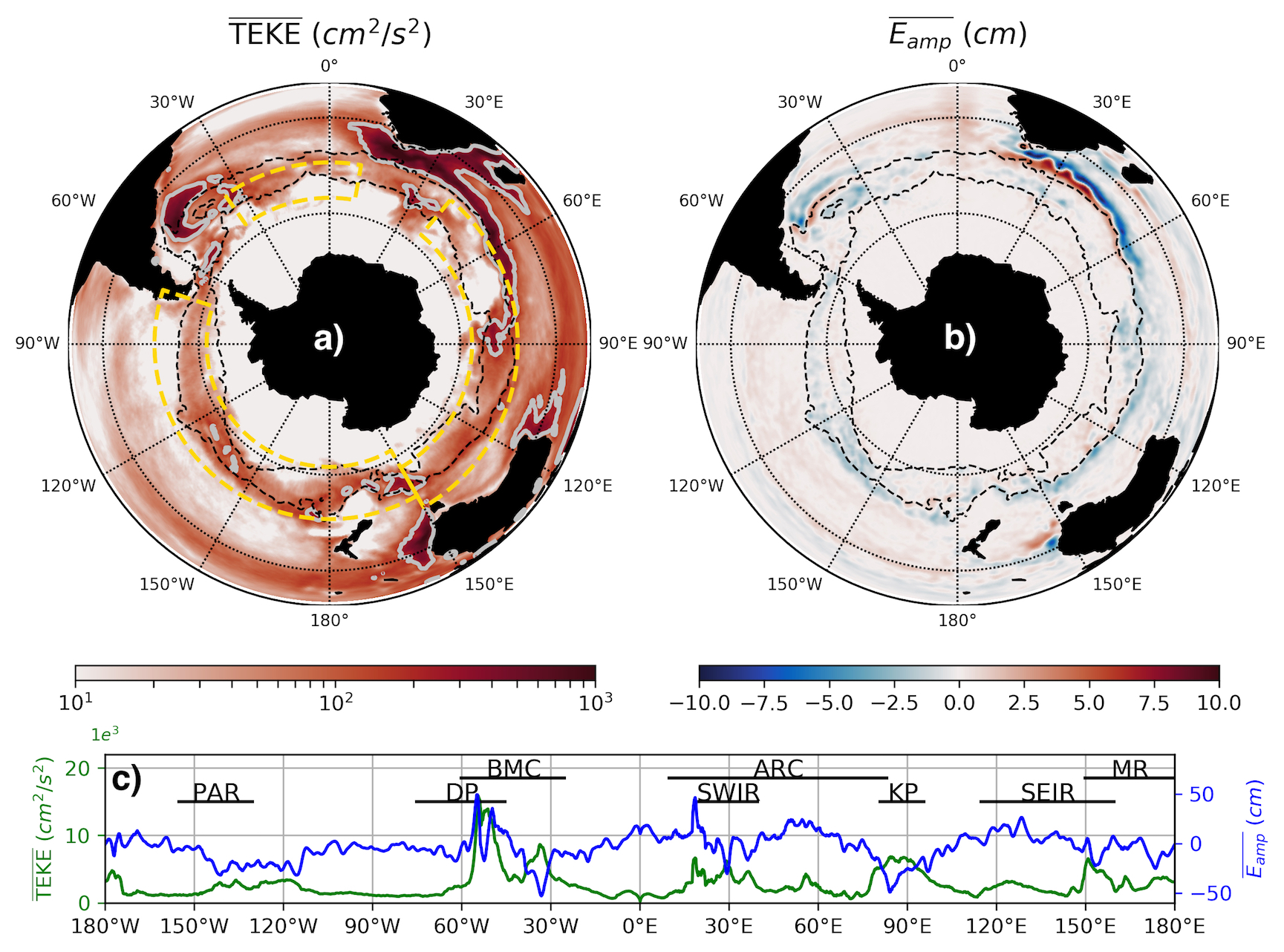}
	\caption{TEKE and mean eddy amplitude maps of the Southern Ocean with mean circumpolar streamlines defining outer edges of the ACC band ($\overline{\textrm{SSH}}$ = $-0.8$ to $0.2$ $m$). 
		a) $\overline{\textrm{TEKE}}$ climatology over the satellite altimetry era from 1993 to 2017. Gray contours correspond to values larger than $183\ cm^2/s^2$.  b) $\overline{E_{amp}}$ or mean eddy amplitude shows areas with high eddy intensity and their polarity dominance. This metric is consistent at the western boundary currents with the deviation of the of sea level (skewness) reported by \citet{Thompson_Skewness_2006}.
		c) Meridional sum of $\overline{\textrm{TEKE}}$ and $\overline{E_{amp}}$ by longitude within the ACC band defined by the black dashed lines in (a,b). Yellow boxes in a) show the ACC Pacific, Indian and Atlantic basins.}
	\label{fig:spatial_structures}
	\end{figure}

	The proposed decomposition is a robust method to separate coherent eddies from the transient field. $\overline{\textrm{TEKE}}$ hotspots are shown as gray contours in figure~\ref{fig:spatial_structures}a; these regions have $\overline{\textrm{TEKE}} \geq 190\  cm^2/s^2$ (2$\sigma$), i.e., more than 3 times the SO average of $\sim44\ cm^2/s^2$. These areas are associated with interactions between the ACC and major bathymetric features, and western boundary currents. The prominent topographic features are the Pacific Antarctic Rise (PAR; 155$^\circ$W - 130$^\circ$W), Drake Passage (DP; 75$^\circ$W - 45$^\circ$W), Southwest Indian Ridge (SWIR; 20$^\circ$E - 40$^\circ$E), Kerguelen Plateau (KP; 81$^\circ$E - 96$^\circ$E), Southeast Indian Ridge (SEIR; 115$^\circ$E - 160$^\circ$E), and Macquarie Ridge (MR; 160$^\circ$E - 180$^\circ$E). The western boundary currents correspond to the Agulhas Return Current (ARC; 10$^\circ$E - 83$^\circ$E) and the Brazil-Malvinas Confluence (BMC; 60$^\circ$W - 25$^\circ$W). These TEKE hotspots have strong eddy activity, and they have been shown to play a key role in the SO exchange of heat and carbon and upwelling pathways \citep{Woloszyn_Testing_2011,Dufour_Role_2015,Foppert_Eddy_2017,Tamsitt_Spiraling_2017}; even studies using sub-mesoscale resolving simulation show the importance of these hotspots in the transient vertical heat transport \citep{Su_Ocean_2018}.
	
	The TEKE hotspots also have a co-located large signature in the mean amplitude of the reconstructed coherent eddy field ($\overline{E_{amp}}$) for the satellite period (figure~\ref{fig:spatial_structures}b), which highlights regions that are dominated by eddies of one polarity. For example, all western boundary currents have a large negative value of $\overline{E_{amp}}$ equatorial-wards and positive $\overline{E_{amp}}$ signature pole-wards of the climatological currents. This signal is a consequence of the meanders becoming unstable and generating cold core eddies on one side of the climatological jet location and warm cores eddies on the other side. Meanwhile, in the Pacific and Atlantic basins there is a positive eddy amplitude signature north of the ACC (dashed lines), while the Indian sector has a negative signature. 
	Figure~\ref{fig:spatial_structures}c further shows the  $\overline{\textrm{TEKE}}$ (green curve) and $\overline{E_{amp}}$ (blue curve) meridionally integrated across the climatological ACC ($\overline{\textrm{SSH}} = $ $-0.8$ to $0.2$ $m$), as well as the major topographic features denoted by horizontal lines. Note that downstream of each of the major topographic features with a TEKE peak there is a change in the polarity of $\overline{E_{amp}}$. In the case of the Pacific Antarctic Rise, Drake Passage, Agulhas Return Current, Southwest Indian Ridge, and Southeast Indian Ridge there is a transition from positive to negative $\overline{E_{amp}}$, while at Kerguelen Plateau and Macquarie Ridge the transition is from negative to positive $\overline{E_{amp}}$. We suspect this transition between coherent eddy polarities is dependent with the generation of eddies through detachment from the meanders. However, this change in polarity is not always located near intense currents and therefore it should be further investigated.
	
	\subsection{Trends}
	
\begin{figure}[!t]
		\centering
		\includegraphics[width=0.8\linewidth]{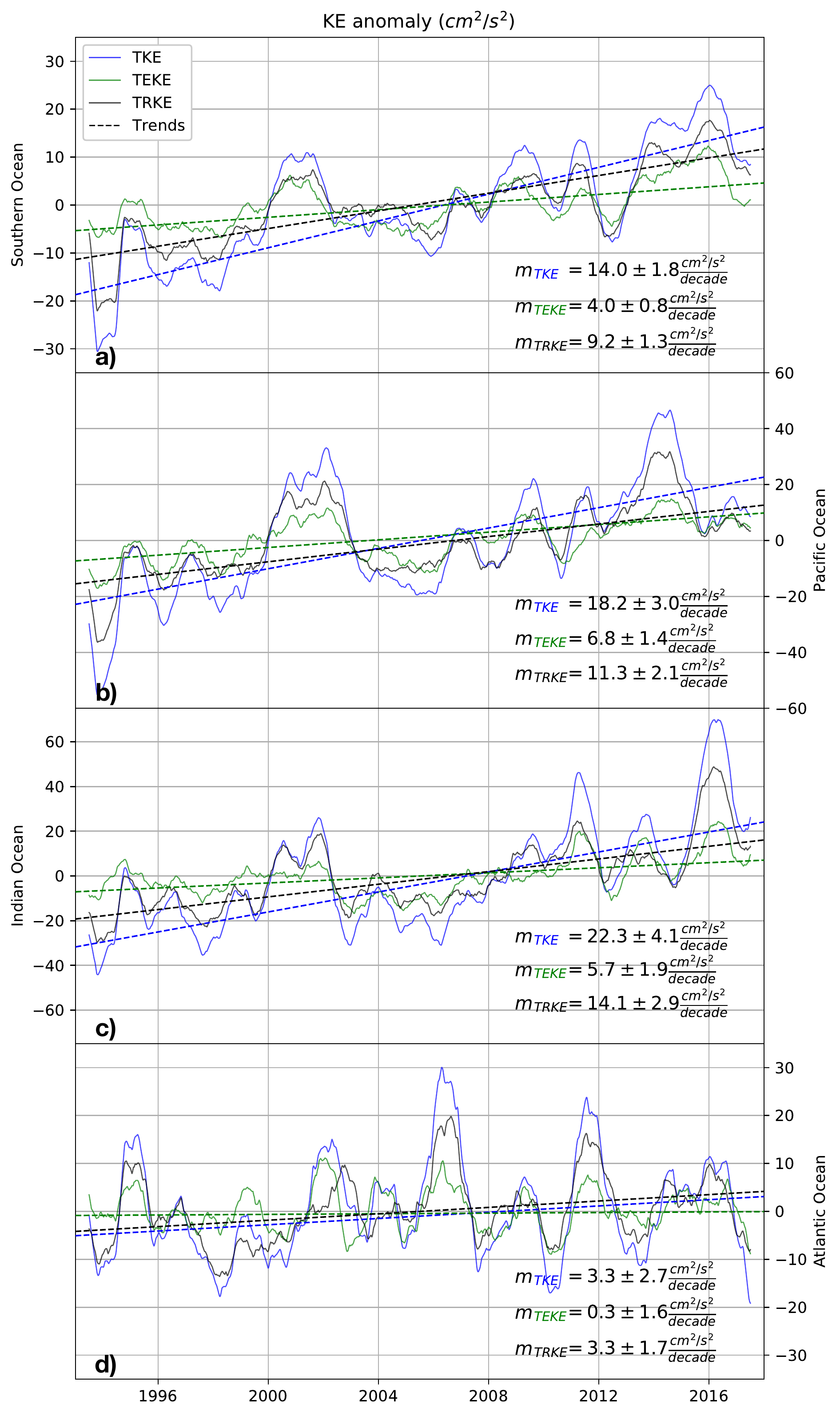}
		\caption{Time series of transient kinetic energy (blue line), transient eddy kinetic energy (green line), and transient residual kinetic energy (black line) anomalies relative the satellite time period 1993-2017 for a) the Southern Ocean (SO) and three SO sectors: b) Pacific Ocean, c) Indian Ocean, and d) Atlantic Ocean. 
		Solid lines show running annual means, while the dashed line shows the 95\% confidence satellite altimetry era trend.}
		\label{fig:trends_SO}
	\end{figure}

	Now we further explore the reported increase of TKE trends over the satellite record \citep{Hogg_Recent_2015}. Figure~\ref{fig:trends_SO} shows time series of the running annual average anomaly of TKE and its decomposition (TEKE and TRKE) spatially averaged over the SO and three sectors in the SO similar to those used by \citet{Meredith_Circumpolar_2006} and  \citet{Hogg_Recent_2015}: SO: 0$^\circ$E - 360$^\circ$E, 30$^\circ$S - 60$^\circ$S; Indian Ocean: 40$^\circ$E - 150$^\circ$E, 44$^\circ$S - 57$^\circ$S; Pacific Ocean: 150$^\circ$E - 288$^\circ$E, 48$^\circ$S - 62$^\circ$S; and Atlantic Ocean: 325$^\circ$E - 10$^\circ$E, 46$^\circ$S - 56$^\circ$S (Dashed yellow boxes in figure~\ref{fig:spatial_structures}a). Dashed lines in figure~\ref{fig:trends_SO} show the linear trends with 95\% confidence. Note that the magnitude of the variability remains constant in time for all basins, and there are no step changes where the number of satellites has increased.  Therefore, we infer the increasing signal is an intrinsic response of the transient field.
	
	The SO energy anomaly magnitude is smaller than that in the Pacific and Indian sectors, as it includes large areas of the South Pacific, South Atlantic, and Indian gyres where the KE content is lower than the other sectors which were selected to mostly cover sections of the ACC. However, significant increasing trends are observed for each KE component (figure~\ref{fig:trends_SO}a \& table~\ref{tbl:trends}).
	The contributions of TEKE and TRKE have the same magnitude and are consistent with the TEKE and TRKE spatial averages: TEKE explains between 30-50\% and TRKE 50-70\% of TKE over the time-series. 
	
	
	The Pacific sector of the SO shows significant increasing trends for the transient kinetic energy and all its components, where the TKE trend is constituted by $34\pm12\%$ of TEKE and $66\pm12\%$ of TRKE. The Indian sector also shows an increasing trend for TKE, TEKE and TRKE and the contribution of TEKE to the TKE trend is $39\pm14\%$, while TRKE is responsible of $61\pm13\%$ of the TKE trend. Meanwhile, the Atlantic sector only shows a significant increase in TKE and TRKE, but TEKE trend is statistically indistinguishable from zero. 
	
	The detected TKE trends found from gridded data using TrackEddy and the geometrical reconstruction of the eddy field are consistent with the trends calculated from satellite tracks by \citet{Hogg_Recent_2015} (Table \ref{tbl:trends}). 
	However, \citet{Hogg_Recent_2015} also noted that TKE trends computed from along satellite tracks are larger than those calculated from gridded data by a factor of 1.9 in the Pacific, 1.7 in the Indian, and 1.6 in the Atlantic. Therefore, even when the detected trends are consistent, they could be still underestimated by the interpolation from tracks to gridded data.
	
	\begin{table}[!t]
		\caption{Detected trends of  TKE by \citet{Hogg_Recent_2015} from satellite tracks,  AVISO+ gridded data and decomposition trends (TEKE, TRKE) over each basin in $cm^2/s^2$ per decade.}
		\label{tbl:trends}
		\begin{center}
			\begin{tabular}{ccccc}
				&  SO  & Pacific & Indian & Atlantic\\
				\hline
				\begin{tabular}{@{}c@{}}TKE \\ (Hogg et al. 2015)\end{tabular}	& \textbf{---} &$14.9\pm4.1$ & $18.3\pm5.1$ &$4.0\pm3.7$\\
				\hline
				TKE		& $14.0 \pm 1.8$ &  $18.2 \pm 3.0$ & $22.3 \pm 4.1$ &$3.3 \pm 2.7$  \\
				TEKE   & $4.0 \pm 0.8$  &  $6.8 \pm 1.4$ & $5.7 \pm 1.9$ & $0.3 \pm 1.6\ $ \\
				TRKE   & $9.2 \pm 1.3$   &  $11.3 \pm 2.1$ & $14.1\pm 2.9$  & $3.3 \pm 1.7$  \\
			\end{tabular}
		\end{center}
		\label{default}
	\end{table}
	
	The increase in the TKE signal is composed mostly of the addition of the TEKE and TRKE trends. Even when TEKE fluctuates between 30 to 50 percent of the TKE signature, it can be attributed uniquely to coherent eddies, while the residual TRKE still includes large scale jets, meanders, wave processes and some misidentified eddies. This decomposition has identified the contribution of mesoscale processes to the observed trend in the SO transient kinetic energy; the adjustment of properties of the coherent eddy field are explored in the following section. 
	
	\subsection{Eddy Characteristics}
	
	The increase in TEKE previously described highlights that part of the observed Southern Ocean TKE trend is due to changes in the coherent eddy field. These results suggest that one or more eddy properties (number, amplitude, area, and/or eccentricity) have increased over the last two decades. We investigated the eddy characteristics responsible for the positive TEKE trends using the individual geometric characteristics of each identified eddy from TrackEddy output. We diagnosed the time series of each of the properties, which include the number of eddies, eddy amplitude, eddy area, eccentricity, and eddy orientation. The variables showing a robust trend were the number of eddies ($E_n$), the absolute eddy amplitude ($|E_a|$), defined as the maximum absolute amplitude within each identified eddy, and the eddy area ($E_{area}$), defined as the area of the region containing the identified closed contour.
	
\begin{figure}[!t]
		\includegraphics[width=1\linewidth]{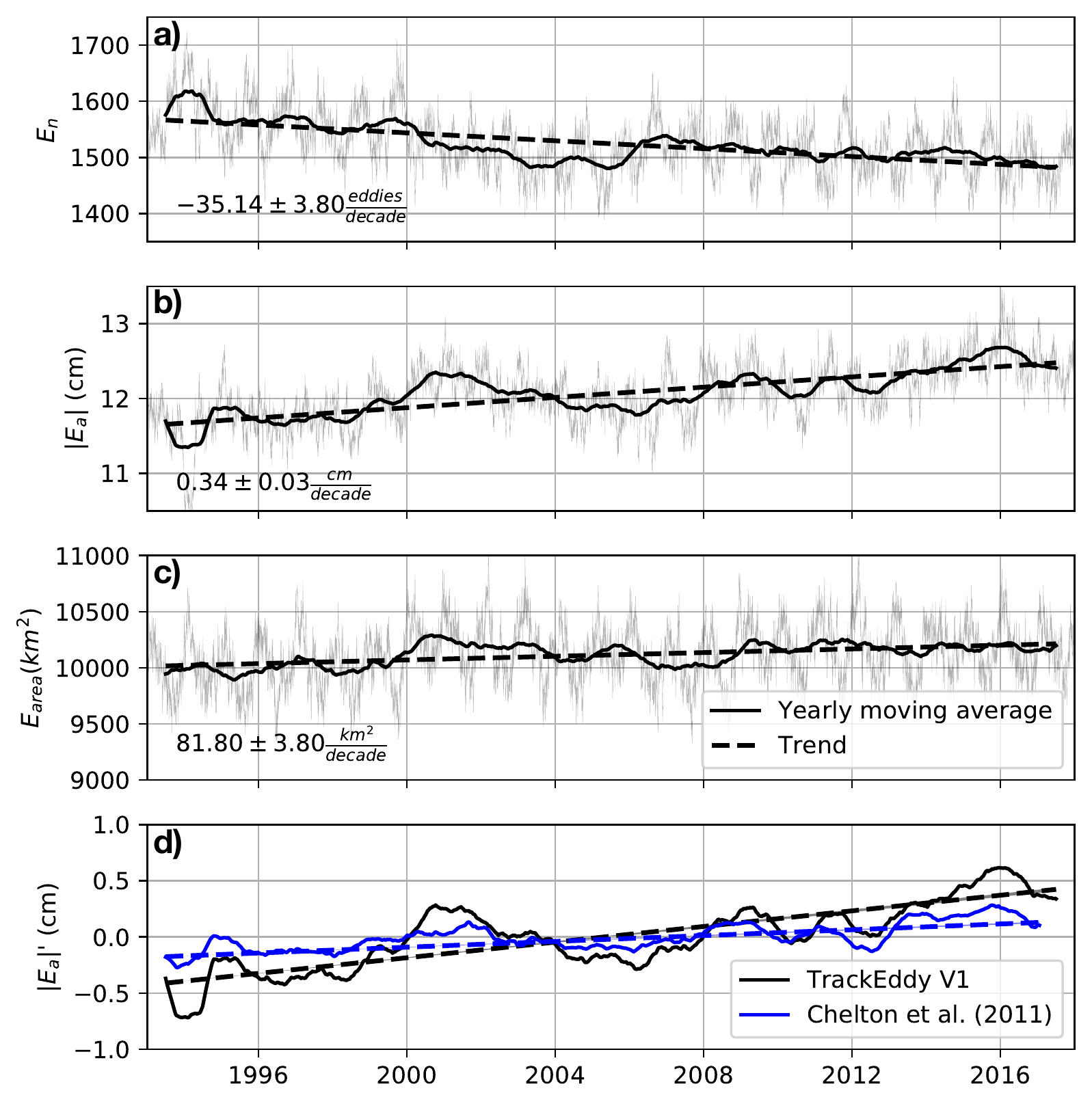}
		\caption{ Time series of  a) the number of detected eddies, b) the eddy mean absolute amplitude and c) the eddy mean area over the Southern Ocean from TrackEddy, and d) the comparison between the normalized TrackEddy mean eddy amplitude time anomaly ($|E_a|'$) and \citet{Chelton_Global_2007}.  
		Colored lines show running annual means, while the dashed line shows the satellite altimetry era trend.}
		\label{fig:eddy_properties_trend}
	\end{figure}

	The average detected number of eddies in the SO over the satellite record is around $1500$ per daily snapshot. Figure~\ref{fig:eddy_properties_trend}a shows daily variability, where the observed seasonal cycle peaked during October is attributed to a lagged response between the eddy field and the seasonality of the mixed layer \citep{Nardelli_Southern_2017} and is consistent with the sub-mesoscale observations presented by \citet{Yu_An_2019}. Additionally, the running annual mean shows a significant decrease of $-35.14$ eddies per decade. This signal is counter-intuitive, as it shows that an increase in TEKE  does not depend on the number of identified coherent mesoscale eddies. We still do not know the mechanism which drives the decrease in the number of eddies, but we believe that understanding the mechanism could be crucial to further understand eddy saturation.
	
	Meanwhile, the mean eddy amplitude and mean eddy area have increased at a rate of $0.34\ cm$ and $81.8\ km^2$ per decade respectively (figures \ref{fig:eddy_properties_trend}b,c). As the relative trend of the eddy amplitude is larger than the relative trend of eddy area, the TEKE trends are mostly explained by the intensification of the eddy amplitude with a small contribution from the eddy area. Note that eddies with a large increase in amplitude and a small increase in area will produce larger SSH gradients and therefore stronger geostrophic velocities.
	The eddy amplitude intensification qualitatively agrees with the trends computed from  the dataset of \citet{Chelton_Global_2007}  (figure~\ref{fig:eddy_properties_trend}d). The mean eddy amplitude variance as computed by TrackEddy is around 10 times larger than results from \citet{Chelton_Global_2007}, and the detected trend by TrackEddy ($0.34\ cm$ per decade) is three times larger than Chelton's ($0.1\ cm$ per decade). This difference is attributed to how the algorithms report the amplitude of eddies. The TrackEddy definition corresponds to the maximum SSHa within the eddy, while Chelton's algorithm uses the maximum SSHa value minus the discrete level in which the eddy was identified and applies a zonal high-pass filter and a half-power filter cutoffs of 3 degrees by 3 degrees over 20 days periods. For example, take an eddy that is identified at the 10~$cm$ closed contour,  with a maximum SSHa elevation of 100~$cm$. TrackEddy defines the amplitude as 100~$cm$, while Chelton's algorithm defines the amplitude as 90~$cm$. As the identification level may change depending on the eddy characteristics, a definition of amplitude dependent on the identified level will reduce the detected signal.
	
\begin{figure}[!t]
		\includegraphics[width=1\linewidth]{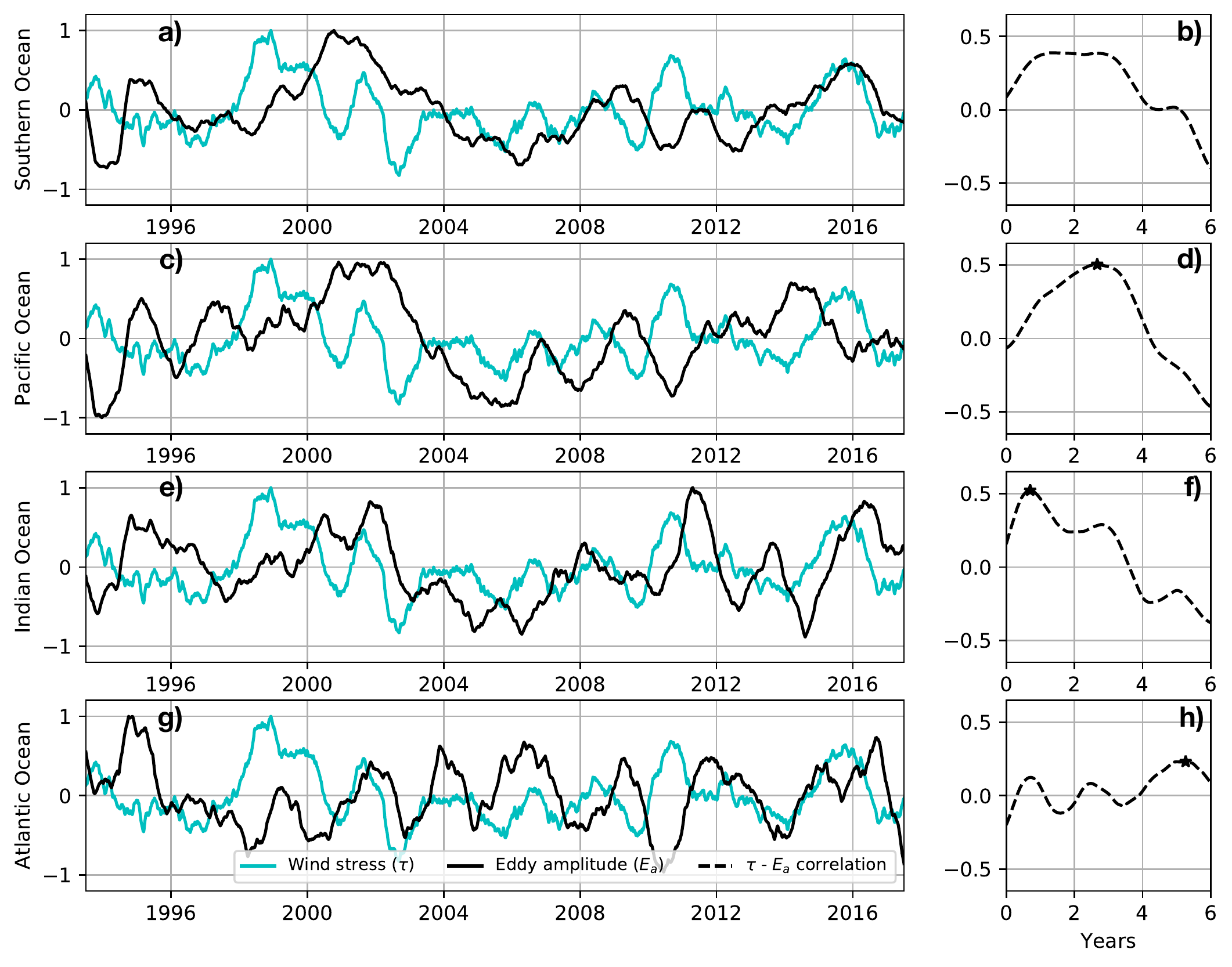}
		\caption{De-trended and normalized time series of the annual running average of eddy mean amplitude (black line) in the a) Southern Ocean, c) Pacific Ocean, e) ndian Ocean, and g) Atlantic Ocean from satellite data and the mean wind stress anomaly from JRA-55-do (cyan line). Dashed lines in  plots b, d, f, and h corresponds to the cross-correlation between lagged winds and the mean eddy amplitude. The maximum absolute correlation is shown by the stars.}
		\label{fig:eddy_lag}
	\end{figure}

	As discussed in the introduction, the transient field has responded to the intensification of the westerly winds in the SO \citep{Swart_Observed_2012,Bracegirdle_Assessment_2013,Lin_Mean_2018,Young_Multiplatform_2019}. Furthermore, previous studies have shown a lagged response between the wind stress and TKE trends \citep{Hogg_Interdecadal_2006,Morrow_Eddy_2010,Patara_Variability_2016}. To explore the inter-annual response of the coherent eddy field and the winds in the SO, we removed the long-term trend and then compute the cross correlations between the de-trended and normalized mean eddy amplitude and the de-trended and normalized mean wind stress calculated using the bulk formula without the ocean state component from JRA55-do \citep{Tsujino_JRA_2018} (figure~\ref{fig:eddy_lag}). 
	
	The SO time series of mean eddy amplitude shows a weak correlation with the SO wind stress (figure~\ref{fig:eddy_lag}a). The lagged cross-correlation of these time series has two predominant maxima from 1 to 3 years (figure~\ref{fig:eddy_lag}b). \citet{Hogg_Interdecadal_2006} suggested that the slow response corresponds to strong topographic steering due to the vertical momentum transport from interfacial form stress of the transient field, while a possible hypothesis to the fast response could be the direct enhancement-readjustment of baroclinic instabilities \citep{Abernathey_Topographic_2014}.
	The Pacific Ocean cross correlation has a clear maximum lag at 3 years (figure~\ref{fig:eddy_lag}c-d), suggesting the response of the eddy field in the Pacific sector is mostly dominated by the topographic steering mechanism. The Indian Ocean has two local maxima in the cross correlation (figure~\ref{fig:eddy_lag}e-f), where the largest peak has a lag of 8 months, again suggesting a fast response of the eddy fields to the winds. Finally, the lagged cross-correlation in the Atlantic Ocean is not significant, however it still shows three maxima at 1, 3, and 5 years (figure~\ref{fig:eddy_lag}g-h).
	
	
	The SO eddy field could be responding the winds through a fast-baroclinic adjustment and a slow interfacial transfer of momentum. Moreover, this response varies in each of the basins, which suggests a spatial dependence possibly related to the main topographic features of the SO basins. 
		
	\section{Discussion and Conclusions}

  	We present here a new eddy-reconstruction algorithm to extract the kinetic energy contained in mesoscale coherent eddies. Our synthetic tests  show that the Transient Eddy Kinetic Energy is well estimated by TrackEddy and the method is sensitive enough to extract  the energy signature contained only by coherent eddies. Taking advantage of the 23 years of the AVISO+ SSH, we identified and reconstructed each eddy based on its geometric parameters: amplitude, area, orientation, and eccentricity. 

	The Transient Eddy Kinetic Energy (TEKE), that is the transient energy contained in coherent eddies,  in the Indian and Pacific sectors of the SO exhibits a significant trend over the satellite altimetry era. 
	Consistent with previous studies \citep{Hogg_Recent_2015}, Transient Kinetic Energy (TKE) trends are explained by a combination of the changes in the eddy and residual fields, where TEKE explains $1/3$ of the TKE while TRKE explains most of the remaining $2/3$. Note that this is still an underestimation of the eddy contribution as TrackEddy does not capture all eddies due to its rigorous criteria. However, it is clear that the contribution of non-coherent processes (TRKE) is crucial to further understand the transient kinetic energy.

	In addition, we find an intriguing decadal increase in the eddy amplitude, and a decrease in eddy numbers in the SO since 1993, which is responsible for most of the increase in TEKE. There is a correlation between the 1-3 year lagged wind stress and the eddy amplitude in the SO, which could be the response of the eddy field to a fast-baroclinic and a slow interfacial form stress mechanism. The largest cross-correlations were found in the Pacific and Indian sectors and they are consistent with the lagged TKE response of 2 to 3 years to the intensification of the SO westerly winds.
	Overall, these results suggests a response of the coherent eddy field to intensification of westerly winds in the SO, and this is consistent with the lag found in previous studies.
	
    Determining changes to the transient eddy field is fundamental to our understanding of the SO and its potential response to climate change. 
    The Antarctic Circumpolar Current (ACC) comprises eddies, jets, and wave processes. Therefore,  understanding the transient variability of the ACC will help us to assess global changes of heat transport \citep{Screen_The_2009} and carbon subduction \citep{Keppler_Regional_2019}.
    The presented results indicate that the SO coherent eddy field may be responding to the climate change signal in the wind stress, and motivates us to achieve a better understanding of each process. This hypothesis will be further explored in more detail as a continuation of this research. 
    
    There is scope for the proposed method to be refined further in future studies.
    First, the active resolution of AVISO+ limits the capabilities of TrackEddy to capture small coherent eddies, whereas future wide-swath satellite altimetry missions (SWOT) will capture all mesoscale and a considerable section of sub-mesoscale processes. Conceptually, we presume that TrackEddy could be implemented to analyze different motion scales and provide a better understanding of interaction between coherent eddies at different scales. 
    Second, the current KE decomposition only provides a simple estimate of the TRKE, which could be further separated into the jet and wave flow components. 
    Third, the estimation of TEKE could be improved by further enhancing the optimization fitting code, which currently relies on fixing some eddy properties to constrain the optimization.
    We suspect that by introducing additional parameters such as vorticity and/or the phase angle between the meridional $v$ and zonal $u$ components, the identification and reconstruction of eddies could be improved.
    Finally, the assumption of Gaussian eddies may well be violated under strong eddy-eddy, eddy-waves, or eddy-jet interactions, therefore a more complex function could be fitted to represent strong interactions.

    In summary, we have developed a new eddy-tracking algorithm with the capability to reconstruct the eddy field and calculate its kinetic energy. 
    We  find that the decadal increase in TKE in the SO since the early 1990s is explained by trends in each mesoscale process (coherent eddies and residual).
    The coherent eddy field has a clear response to the winds intensification and therefore to climate change. This response may have implications for the efficiency of carbon and heat sinks in the Southern Ocean.
	
	%
	%
	%
	%
	%
	%
	%

	\acknowledgments
	The satellite altimeter products were produced by Ssalto/Duacs and distributed by AVISO+, with support from CNES (\url{https://www.aviso.altimetry.fr/en/data/products/ sea-surface-height-products/global/gridded-sea-level-heights-and-derived-variables.html}).
	The identified eddies are available in CSV, parquet, HDF5 and netCDF format at \url{http://dx.doi.org/10.25914/5cb6859e4df3e}. Some examples on how to load and post-process the TrackEddy output can be found at the TrackEddy GitHub repository (\url{https://github.com/josuemtzmo/trackeddy}).
	Murray Fang provided constructive criticism on the first draft of this manuscript. J.M-M. was supported by the Consejo Nacional de Ciencia y Tecnolog\'{i}a (CONACYT), Mexico funding. A.K.M. was supported by the Australian Research Council DECRA Fellowship DE170100184. Computations were undertaken on the National Computational Infrastructure (NCI) in Canberra, Australia, which is supported by the Australian Commonwealth Government.

	
	
	
	

\end{document}